\documentstyle[11pt]{article}
\def\Tr{ \tilde{\rm Tr}\,} 
\newcommand{\IR}{\relax{\rm I\kern-.08em R}} 
\newcommand{\IZ}{\relax\ifmmode\mathchoice 
{\hbox{\cmss Z\kern-.4em Z}}{\hbox{\cmss Z\kern-.4em Z}} 
{\lower.9pt\hbox{\cmsss Z\kern-.4em Z}} 
{\lower1.2pt\hbox{\cmsss Z\kern-.4em Z}}\else{\cmss Z\kern-.4em 
Z}\fi} 
\newcommand{\beq}{\begin{equation}} 
\newcommand{\eeq}{\end{equation}} 
\newcommand{\beqn}{\begin{eqnarray}} 
\newcommand{\eeqn}{\end{eqnarray}}

\newcommand{\N}{{\scriptscriptstyle N}}

\begin{document} 
\begin{titlepage} 
\title{\vskip -60pt 
{\small 
\begin{flushright} 
KIAS-P00075\\
UOSTP-00108\\ 
{\tt hep-th/0011244} 
\end{flushright}} 
\vskip 45pt 
Comments on Noncommutative Gauge Theories\\\ 
~} 
\vspace{4.0cm} 
\author{\\ 
\\ 
\\ 
Dongsu Bak,${}^{a}{}^{\natural}$\,\,  
Kimyeong Lee${}^{b}{}^{\dagger}$ and Jeong-Hyuck Park${}^{b}{}^{\ddagger}$} 
\date{}
\maketitle 
\begin{center} 
\textit{${}^{a}$Physics Department, University of Seoul, Seoul 
130-743, Korea}\\ 
~\\ 
\textit{${}^{b}$School of Physics, Korea Institute for Advanced Study}\\
\textit{207-43 Cheongryangri-dong Dongdaemun-gu, Seoul 130-012, Korea}
\end{center} 
\vspace{1.0cm} 
\begin{abstract} 
\noindent We study the gauge theories on noncommutative 
space. We employ the idea of the covariant position to understand the 
linear and angular momenta, the center of mass position, and to 
express all gauge invariant observables including the Wilson line. In 
addition, we utilize the universality of the $U(1)$ gauge theory, 
which originates from the underlying matrix theory, to analyze various 
solitons on $U(N)$ theories, like the unstable static vortex solutions 
in two dimensions and BPS dyonic fluxon solutions.
\end{abstract} 

\vskip 3cm

$\overline{\mbox
{ Electric~\,Correspondence:}~~{}^{\natural}\mbox{dsbak@mach.uos.ac.kr},~\,{}^{\dagger}\mbox{klee@kias.re.kr},~\,{}^{\ddagger}\mbox{jhp@kias.re.kr}~~~}$ 
\thispagestyle{empty} 
\end{titlepage} 
\newpage 
\setcounter{footnote}{0} 
 
\section{Introduction} 
Recently the classical properties of noncommutative gauge theories
have been investigated in many
directions\cite{aschwarz}-\cite{terashima}.
 There are several 
interesting issues raised recently on the noncommutative gauge 
theories. 
First issue is about the gauge invariant observables. Gross, Hashimoto
and Itzhaki showed that for any gauge covariant local quantity, one
can construct the gauge invariant quantity by taking the trace over
space with the open Wilson line\cite{itzhaki}.  Second concerns about
the $U(1)$ universality as noted by Gross and
Nekrasov\cite{nekrasova}. It seems that it is possible to everything
about the $U(N)$ gauge theory can be found in the $U(1)$ theory.
 
In this letter, we elaborate these issues. We introduce a {\it
covariant position operator} and show that the gauge invariant
operators constructed in Ref.~\cite{itzhaki} are the ``Fourier
transformation'' of a given operator with respect to the covariant
position operator. Then we elaborate the $U(1)$ universality and argue
that the underlying matrix theory is the reason behind it. Finally, we
explore this universality in some concrete examples. The physics of
$M$ D0 branes on $N$ D2 branes is explored.  Also we provide the
solution of $M$ fluxons in the $U(N)$ theory.  Our discussion will be
about the classical aspect of the noncommutative gauge theory.
 
Some of points made in this paper are not quite new, but presented as
they illuminate and emphasize the different aspect of the similar
ideas. The Wilson line observables have been introduced by Ishibash,
Iso, Kawai, and Kitazawa before\cite{iso}.  GHI elaborated this idea
and extended to include other operators\cite{itzhaki}.  In this paper,
we short-circuit this idea by introducing the covariant position
operator $X_i = x_i + \theta \epsilon_{ij} A_j $, which is related to
the covariant differential operators $D_j$ by $X_i =i\epsilon_{ij}
\theta D_j$.  Then, the observables constructed by IIKK and GHI are
just the Fourier transformation of a given local covariant operator
with the covariant position operator.  It is much simpler to think
this way. Especially one can define even a delta function on
noncommutative space, which would lead to almost localized gauge
invariant observables. 
 
The idea of the gauge covariant position operator appears also 
naturally when one considers the conserved linear and angular 
momenta, the center of mass, etc. The gauge invariant angular 
momentum density in the standard field theory is $\epsilon_{ij} x_i 
T_{0j}$. One may wonder what is the corresponding formula in the 
noncommutative gauge theory as the energy momentum tensor is just 
gauge covariant, not invariant. The gauge covariant position operator 
appears naturally instead of  ordinary position operators. Similarly, 
the center of the mass position can be defined with the covariant 
position operators.  When there is no matter in the fundamental 
representation, the translation and rotational transformations of 
matter and gauge fields are gauge equivalent to just a transformation 
of the gauge field. This observation allows to reexpress the linear 
and angular momenta in only  the gauge field or the covariant position. 
 
GN have shown that the $U(1)$ gauge theory on the noncommutative plane
with adjoint matters only has many gauge nonequivalent vacua
characterized by a natural number $N$, and have argued that they
correspond to the vacua of the $U(N)$ gauge theory\cite{nekrasova}.
In addition, they argue that all $U(N)$ theories are a part of the
$U(1)$ theory.  There is a good reason behind this universality.  We
elaborate this point of view more carefully. Basically the degrees of
freedom do not change, but one can divide, or regroup the degrees of
freedom living on a space into some on a new space and some on
internal space. It is a sort of regrouping the degrees of freedom of
the $U(1)$ theory. We elaborate this in some detail. Of course the
underlying reason for this is that there is a hidden matrix mechanics
model, which leads to all $U(N)$ theories. The covariant position
operators discussed above are those appear in the matrix theory
naturally.
 
Due to this universality, one can write the solutions of the $U(N)$ 
gauge theories in the $U(1)$ theory more conveniently. We find the 
$M$ static magnetic flux in the $U(N)$ theory. This can be interpreted as 
a $M$ D0 branes on $N$ D2 brane background. We find the mass of the 
$M\times N$ tachyonic modes around this solution. The fluxon solution on the 
$U(1)$ theory is a composite of two BPS monopoles of opposite magnetic 
charge. We find the $M$ dyonic fluxon solutions in the $U(N)$ gauge 
theory which are 1/2 BPS.

The plan of this paper is as follows: In Section 2,  we review the 
noncommutative gauge theory and introduce the covariant position 
operator. In Section 3, we review the $U(1)$ universality. In Section 
4, we provide the solutions of the field equation in two and three 
dimensions.

\section{The Covariant Position Operator}

We consider the gauge theory on noncommutative plane whose
coordinates are $(x,y)$.  The coordinates $x,y$ satisfy  the relation
\begin{equation} 
[x,y] = i \theta 
\label{comm} 
\end{equation} 
with $\theta>0$. This noncommutative plane has not only the 
translation symmetry but also the rotational symmetry. One can see that 
the parity operation $(x,y)\rightarrow (x,-y)$ is broken on 
noncommutative plane. The classical field on this noncommutative space 
is an element $\Phi(t,x,y)$ in the algebra ${\cal A}_\theta$ defined 
by $x,\, y$ with the relation $(\ref{comm})$.

Let us introduce the complex coordinates of the noncommutative space, 
\beq 
c= \frac{1}{\sqrt{2\theta}}(x+iy), \;\;~~~ \bar{c} = 
\frac{1}{\sqrt{2\theta}}(x-iy) \, ,  
\eeq 
which satisfy $ [c,\bar{c}] = 1 $.  This commutation relation is that 
of the creation and annihilation operators for a simple harmonic 
oscillator and so one may use the simple harmonic oscillator Hilbert 
space ${\cal H}$ as a representation of (1).  The ground state is 
$|0\rangle$ such that $c|0\rangle =0$,  
and $|n \rangle$  is the usual number eigenstate. 
The integration over noncommutative two plane becomes the trace over 
its Hilbert space, which respects the translation symmetry: 
\beq 
\int d^2 x {\cal O}(x)  \rightarrow \Tr {\cal 
O}(x) \equiv 2\pi \theta  
\sum_{n\ge 0}  {\langle}n|{\cal O} |n \rangle \, . 
\eeq 

We consider mostly the gauge theory with the matter fields
$\phi(t,x,y)$ in the adjoint representation. The gauge fields on the
space are $A_\mu(t,x,y)$. We denote a matter field in the fundamental
representation $\psi$. Because of the commutation relation, the
derivative along the noncommutative coordinate of a function becomes $
\partial_i \phi = [\hat{\partial}_i,\phi]$ with
\beq 
\hat{\partial}_i = \frac{i}{\theta}\epsilon_{ij} x_{j}  \, ,
\eeq 
where $\epsilon_{12}=1$, $x^1=x$ and $ x^2=y$.  For the covariant 
derivatives of the matter fields are defined as $ \nabla_i \phi = 
[D_i, \phi]$, and $ \nabla_i \psi = D_i \psi - \psi \hat{\partial}_i 
$,  where  
\beq 
D_i = \hat{\partial}_i - iA_i = \frac{i}{\theta}\epsilon_{ij} x_j
-iA_i \,  .
\eeq 
The field strength on the noncommutative space is given by $ 
F_{\mu\nu} = \partial_\mu A_\nu-\partial_\nu A_\mu-i[A_\mu,A_\nu]$.  
The local gauge symmetry is given by a unitary transformation $U(x)$ 
$ \psi \rightarrow U\psi $, $ \phi \rightarrow U\phi U^\dagger $, and  
$  A_i \rightarrow UA_i U^\dagger -i [\hat{\partial}_i ,U]U^\dagger $.  
Under this local gauge transformation, the operator $D_i$ transforms 
covariantly, and so are the covariant derivatives of the fields. 
 
The Lagrangian for the gauge theory with the adjoint matter is  
\beq  
L = \frac{1}{e^2}  \Tr \Bigl\{ -\textstyle{\frac{1}{4}} 
F_{\mu\nu} F^{\mu\nu}- \textstyle{\frac{1}{2}}\nabla_\mu\phi^I
\nabla^\mu \phi^I + \textstyle{\frac{1}{4}} \sum_{I,J}
[\phi^I,\phi^J]^2 \Bigr\} \, .  
\eeq 
While one can choose the matter potential arbitrarily, we choose the
theory so that it is a dimensional reduction of the ten dimensional
Yang-Mills theory.   The gauge field $A_0$ is the Lagrangian multiplier
and implies the Gauss law constraint on the initial data,
\beq 
[D_i, E_i] - i [\phi^I, \nabla_0\phi^I] = 0 \, ,
\eeq 
where $E_i=F_{0i}=i\nabla_{0}D_{i}$.  
The local gauge 
transformation $U = e^{i\Lambda}$ is indeed an $U(\infty)$ operator on 
the Hilbert space as 
$\Lambda(x)=\sum_{mn} \Lambda_{mn} |m \rangle   {\langle}n|$.

This theory has several gauge invariant observables.  When there is no
matter field in the fundamental representation, we take the trace over
all Hilbert space to get the gauge invariant operators. IIKK
introduced the gauge invariant observables by using the open Wilson
line\cite{iso}.  GHI found some generalization by incorporating other
covariant quantities\cite{itzhaki}.

In this paper, we give a new spin to this idea by introducing a gauge 
covariant position operator 
\beq 
X_i =  i \theta \epsilon_{ij}D_j = x_i +\theta \epsilon_{ij} A_j  \, .
\eeq 
Clearly in the commutative space limit $\theta=0$, it is a position
operator. We will also see that this operator is the natural operator
from the matrix theory point of view. This position operator can be used to
identify the position of a given soliton.  
As   $B=F_{12}=  -{1\over
\theta}(1+{i\over\theta}[X^1,X^2])$ and
$E_i=F_{0i}= -{1\over \theta} \epsilon_{ij}(\dot{X}^j
-i[A_0,X^j])$,  the Lagrangian can be written completely
with covariant quantities $X^i$ and $\phi^I$.  This is one of the most
striking aspects of the noncommutative gauge theory.

The covariant open straight Wilson  line, which was introduced by IIKK , 
can be put  as  
\beq 
W(\alpha) = e^{-\alpha l^i \hat{\partial}_i} P\exp(i\int_0^\alpha 
 A_i(x+ \beta l) l^i d\beta )   \, ,
\eeq 
which transforms under the local gauge transformation as 
$W(\alpha)\rightarrow U(x) W(\alpha) U(x)^\dagger$. From the 
differential equation satisfied by $W(\alpha)$ and its initial value 
$W(0)$,  we see that it is  identical to   
\beq 
W(\alpha) = e^{-\alpha l^i D_i} = e^{ ip_i X^i} \, , 
\eeq 
where $p_i =- \alpha \epsilon_{ij}l^j/\theta$.  The trace of the 
Wilson line with any covariant operator leads to the gauge invariant 
observable. In our words, it is basically then the Fourier 
transformation of the covariant local operator ${\cal O}(x)$ in terms of the 
covariant position operator, 
\beq 
{\cal O}(p) = \Tr  \left( e^{ip_i X^i} {\cal O}(x) \right) \, .
\eeq 

In our theory any covariant local operator ${\cal O }(x)$ would be a
function of $X^i(x)$ and $\phi^I(x)$ and so ${\cal
O}(X^i(x),\phi^I(x))$. Any function $f(X^i,\phi^I)$ is covariant and
so one can define the weighted invariant quantities ${\cal O}_f = \Tr
f(x) {\cal O}(x) $.
This provides a much broader class of gauge invariant observables. For
example, we define a covariant delta function on the noncommutative
plane as
\beq 
\delta^2(X^i-q^i) = \int \frac{d^2p}{(2\pi)^2} e^{ip_i (X^i-q^i)} \, .
\eeq 
Then one can find a commutative number value by $ \Tr \delta^{2}(X-q)
{\cal O}(x) $.  Due to the operator ordering choice there can be
several definitions of the delta function which are equivalent in the
commutative limit $\theta=0$. Thus, the above quantity does not
measure the localized quantity on the noncommutative space, but a sort
of average over a cell of area size $2\pi \theta$ around the point
$q$. For the fundamental matter field $\psi$, we can find a gauge
invariant operators like $ \bar{\psi}(x) {\cal O}(x) \psi(x) $, which
needs no trace over space.

Among the gauge invariant observables, some of the most prominent ones 
are those related to the symmetries. From the Noether theorem, one can 
find the conserved quantity for each symmetry. 
The time translation symmetry leads 
to the conserved energy  $H=\Tr T_{00}$ to be 
\beq 
H = \frac{1}{2e^2} \Tr \biggl\{ E_i^2 + B^2 + (\nabla_0\phi^I)^2 +(\nabla_i 
\phi^I)^2 - \sum_{I<J}[\phi^I,\phi^J ]^2 \biggr\} \, .
\eeq 
The space translation symmetry ${\cal T}= e^{-l^i \hat{\partial}_i}$, 
which transforms $\phi^I \rightarrow {\cal T}\phi^I  \bar{{\cal T}}$, 
$A_i \rightarrow {\cal T}A_i \bar{{\cal T}}$, also leads to the 
conserved linear momentum $P_i =\Tr T_{0i}$ to be
\beq 
P_i = \Tr \biggl\{ \epsilon_{ij} E_j B + \nabla_0 \phi^I \nabla_i \phi^I 
\biggr\} \, .
\eeq 
As noted by many, the translation  on noncommutative 
plane is gauge equivalent to the shift of the gauge field only, $\phi^I 
\rightarrow \phi^I$, and $A_i \rightarrow A_i - \epsilon_{ij} l^j/\theta$ 
when only adjoint matters are present. This symmetry leads to the 
conserved quantity 
\beq 
P_i =  - \frac{1}{\theta^2} \Tr \nabla_0 X_i \, ,
\eeq 
which is identical to the previous one modulo the Gauss law.

The rotational symmetry leads to the conserved angular
momentum. However, here the subtlety appears for the noncommutative
space. Under the infinitesimal rotational transformation $\delta
\phi^I =\epsilon_{jk}x_{j}[\hat{\partial}_k,\phi^I]$ and $\delta A_i
=\epsilon_{jk}x_{j}[\hat{\partial}_k,A_i] + \epsilon_{ij}A_j $, the
Noether theorem leads to the conserved angular momentum $
J=\epsilon_{ij} \Tr X^i T_{0j}$ to be
\beq 
J =\epsilon_{ij} \Tr  X_i \left(  \epsilon_{jk}E_k B + \nabla_0 
\phi^I \nabla_j \phi^I \right) \, .
\eeq 
Here we have written the density in terms of the gauge covariant quantities, 
discarding the boundary terms (or the commutator terms). Note that the 
covariant position operator appears in this expression.  
Again the rotational symmetry is gauge equivalent to the 
transformation of the gauge field only, $\delta A_i = \frac{1}{\theta}X_i $. 
This  leads to the conserved angular momentum 
\beq 
J = - \frac{1}{\theta^2} \epsilon_{ij} \Tr  X_i \nabla_0 X_j  \, ,
\eeq 
which is identical to the previous one modulo the Gauss law.

For localized field configurations, with the covariantly conserved 
energy $\cal E$, not only we can define their energy, but also we can 
define the moments of the covariant positions\cite{park}. For example,
the  center of mass position is defined as 
\beq 
R^i_{\rm cm } = ({ \Tr X^i {\cal E}})/({\Tr {\cal E}})\,. 
\eeq 
The moment of inertia would be 
%
$I = \Tr  X^i X^i {\cal E}$. 
%

\section{The U(1) Universality}

One can easily generalize the $U(1)$ gauge theory Lagrangian to the 
$U(N)$ theory Lagrangian with the gauge field $A_{\N i}$ and adjoint 
matter field $\phi_{\N}$, which are $N$ by $N$ hermitian matrices. The 
spatial variable for the $U(N)$ theory to be $x_{\N}, y_{\N}$. Gross and 
Nekrasov noticed that this theory is a part of the $U(1)$ theory 
studied before. In this section, we elaborate this in detail and see   
the origin of this be  the well known matrix theory. The key reason 
behind this universality is that the Lagrangian can be written in 
terms of covariant  $X^i$ and $\phi^I$.  
 
To start, let us  find the vacua of the  $U(1)$ gauge theory on the 
noncommutative space. We rewrite the gauge field $A_i$ in the 
the complex coordinate 
\beq 
A\equiv A_1 +iA_2 = -i\sqrt{\frac{2}{\theta}}(c-C)\, ,
\eeq 
where $C= (X^1+iX^2)/\sqrt{2\theta}$. The magnetic field is then 
$B=([C,\bar{C}]-1)/\theta$. The vacua of the theory will have the 
zero field strength. Especially, the magnetic field should vanish. The 
most general solution of this constraint has been found. To write that, 
we  regroup the states in the Hilbert space for any natural number $N$ 
as  
\beq 
|p,\alpha \rangle   = |pN +\alpha\rangle   \, ,
\eeq 
where $p=0,1,2,...$ and $\alpha = 0,1,...,N-1$. The  general 
solution of the zero magnetic field strength modulo gauge 
transformations  is  
%
$C = c_{\N}$, 
%
where $c_{\N}$ and $\bar{c}_{\N}$ are annihilation and creation operators on 
index $p$ so that 
\beq 
c_{\N} =\sum_{\alpha=0}^{N-1}{}\sum_{p=0}^\infty\sqrt{p+1}|p,\alpha \rangle
{\langle}p+1,\alpha|  \, .
\eeq 
The natural number  $N$ 
counts the dimension of the kernel of the operator $\bar{C} C$. (This 
is also  noted by Gross and Nekrasov recently\cite{nekrasova}.)  
The vacuum gauge field is 
then 
\beq 
A = -i\sqrt{\frac{2}{\theta}} (c-c_{\N}) \, .
\label{nvac}
\eeq 
Thus the natural number $N$ denotes the gauge inequivalent vacua. To 
understand the meaning of $N$, we  reexpress all the covariant 
quantities, $D_i$, $X^i$, $B$, $E_i$, $\phi^I$ and $A_{0}$ as  $U(N)$
quantities. For example, 
\beq 
\phi^I =\sum_{mn} 
\phi^I_{mn}|m \rangle   {\langle}n| =\sum_{\alpha\beta}\sum_{pq}
\phi^I_{p\alpha\,q\beta} |p,\alpha \rangle   {\langle}q,\beta|\,  . 
\label{mapping}
\eeq 
Now we introduce the complete set of the $N$ by $N$ hermitian matrices
$T^a,\,a=1,...,N^2$, which are generators of the $U(N)$ group and
express all the coefficient of the quantities in terms of these
matrices, e.g.  $\phi_{p\alpha\,q\beta} = \sum_{a}\phi_{pq}^a
T^a_{\alpha\beta}$.  Also we reexpress the local $U(1)$ gauge
transformation, $U=e^{i\Lambda(x,y)}$, as $\Lambda = \sum
\Lambda^a_{pq} T^a_{\alpha \beta}|p,\alpha \rangle{\langle}q,\beta| $.
When we regard $p,q$ indices as those for the noncommutative space and
$\alpha,\beta$ indices as the internal $U(N)$ gauge symmetry, every
covariant expression of the $U(1)$ theory quantity is reexpressed as one of the
$U(N)$ gauge theory. However the noncommutative coordinates $x_{\N},
y_{\N}$ of the $U(N)$ theory should change only $p,q$ indices not
$\alpha$ and $\beta$ indices.  With $x_{\N},y_{\N}$ defined by 
the $U(N)$ theory are then
\beqn 
x_{\N} = \sqrt{\frac{\theta}{2}} (c_{\N} +\bar{c}_{\N})\,,\ \ \  
y_{\N}= -i \sqrt{\frac{\theta}{2}} (c_{\N}-\bar{c}_{\N})\,, 
\eeqn 
we have $[x_{\N},y_{\N}]=i\theta$.  Letting the covariant position
operator $X^i$ identical in both theories, we define the gauge field
$A_{\N i}$ of the $U(N)$ theory as
\beq 
A_{\N i} = \frac{1}{\theta}\epsilon_{ij}(x_{\N}^j-X^j) \, .
\eeq 
The $N$-th vacuum (\ref{nvac}) of the $U(1)$ theory becomes the
trivial vacuum in the $U(N)$ theory, indicating that the $N$-th vacuum
is indeed the conventional vacuum of the $U(N)$ theory.  Its gauge
field strength $F_{\N 12}$ is covariant and so is identical to that from
the $U(1)$ theory.

Putting them in the Lagrangian and taking the trace over $|\alpha
\rangle {\langle}\beta|$ states we end up with trace over $|p \rangle
{\langle}q|$, which we call $\Tr_{\!\!\N}$ and the $N\times N$ matrix
trace ${\rm tr}_{\N}$.  Then the Lagrangian becomes
\beq L = \frac{1}{e^2}  \Tr_{\!\!\N} {\rm tr}_{\N} 
\Bigl\{ -\textstyle{\frac{1}{4}}   
F_{\N\mu\nu}F_{\N}^{\mu\nu} -\textstyle{\frac{1}{2}}
\nabla_\mu\phi_{\N}^I \nabla^\mu \phi_{\N}^I + 
\textstyle{\frac{1}{4}} \sum_{I,J}  
[\phi_{\N}^I,\phi_{\N}^J]^2 \Bigr\} \, . 
\eeq 
Thus, the $U(N)$ theory is a part of the $U(1)$ theory.

This universality comes about as we regroup the degrees of freedom on
the $x-y$ plane of the $U(1)$ theory to those for the
${x_{\N}-y_{\N}}$ plane of the $U(N)$ theory and those for the
internal space.  This regrouping goes both ways. We can reverse the
argument given here to construct the $U(1)$ theory out of the $U(N)$
theory. Note that the universality works only for gauge theories.

The above universality still holds when there are additional spatial 
dimensions. One needs only one noncommutative plane for such 
regrouping. If there are more noncommutative planes, there would be 
many nonequivalent regrouping of the degrees of freedom.

The $U(N)$ gauge theory on two dimensional noncommutative space is
supposed to describe the $N$ D2 brane dynamics with the background
$B_{12}$ field in the field theoretic limit~\cite{nsw}. Thus, this
universality seems to be mysterious. However there is a underlying
matrix theory where $D0$ branes are basic constituents and $D2$ branes
are composite~\cite{tfss, grt}.  The underlying reason for the
universality is that there is a matrix theory behind all $U(N)$
theories. This mechanics model is a theory with $U(\infty)$
symmetry. The bosonic part of the theory is
\beq 
L = \frac{1}{2g_s} {\rm Tr} \Bigl\{ \sum_{M=1}^9 (\dot{{\cal X}}^M 
-i[{\cal X}^0,{\cal X}^M])^2 + \sum_{M<N}[{\cal X}^M,{\cal X}^N]^2
\Bigr\} \, , 
\eeq 
where $M,N=1,...,9$ and ${\rm Tr}$ is the ordinary trace.

With the coupling $g_s = e^2/(2\pi\theta)$, the dynamics around a
single D2 brane can obtained by ${\cal X}^i = \frac{1}{\theta} x^i +
\epsilon_{ij} A_j = \frac{1}{\theta} X^i~(i=1,2)$, ${\cal X}^{I+2} =
\phi^I~(I=1,2,..7)$, and ${\cal X}^0 = A_0$.  The matrix Lagrangian
becomes the $U(1)$ field theory Lagrangian above plus an infinite
constant term and a topological term.  This matrix theory has also the
solution for $N$ D2 branes and the dynamics around this background is
the noncommutative $U(N)$ gauge theory. Our universality is then the
manifestation of the various solutions of the matrix theory. In
general for the noncommutative gauge theories in $d$-dimensions,
$d>3$, there exists a corresponding $U(\infty)$ underlying field
theory in $d-2$ dimensions and so the universality holds.

The universality is somewhat larger than what we have described here, 
as discussed by  by Aganagic et
al.\cite{aganagic}.  They found the solution which describes the D0
branes not lying on the D2 branes, but separated from the D2
branes. We will see this in more detail in the next section. 
While this is clear in  the matrix theory, it is a something
remarkable in  the field theory.

When there are fundamental matter fields, there is no relation between
the matrix theory and the noncommutative field theory. While the
$U(1)$ theory with fundamental matters can be still rewritten in terms
of the $U(N)$ theory language i.e. the $N\times N$ matrices, there
appears a subtlety to extract a constant from the potential as well as
the sense of the regrouping of the degrees of freedom on space does
not work fully.  In the broken Higgs phase where the matter field has
nonzero expectation value, all $N$ vacua with $N\ge 2$ have constant
energy density when mapped from the $U(1)$ theory.

\nopagebreak 
\section{Examples}

There are several examples of the universality. What is convenient 
about the universality is that one can write the solution of $U(N)$ 
gauge theory in terms of the $U(1)$ theory variables.  Also, the
noncommutative $U(1)$ gauge theory on two dimensions describes the
dynamics of the D0 branes  separated from the D2 branes. This is
rather remarkable from the field theory point of view.

For two dimensional system, there exists a static but nontrivial
magnetic flux solution, generalizing the solutions found in
Ref.~\cite{aganagic}. (Similarly, one may generalize the exact U(1)
vortex solutions\cite{park} to $U(N)$ solutions.)  This can be
interpreted as the $M$ D0 branes in the background of $N$ D2
branes. The general solution is
\beqn 
&& C = \sum_{l=0}^{M-1} \lambda_l |l \rangle   {\langle}l| + \sum_{\alpha,p} 
\sqrt{p+1}|pN+\alpha +M \rangle   {\langle}(p+1)N+\alpha +M|\, , \\ 
&& \phi^I = \sum_{l=0}^{M-1} \varphi_l^I |l \rangle   {\langle}l|
+\sum_{\alpha,p}h_\alpha^I  |pN+\alpha+M \rangle   {\langle}pN +
\alpha+M|  \, . 
\eeqn 
The values $\varphi_l^I$ denote the position of the $l$-th D0 brane
along the transverse coordinates and $h_\alpha^I$ denotes the position
of the $\alpha$-th D2 brane along the same coordinate.  The parameters
$\lambda_l$ are the coordinates of D0 branes on the noncommutative
$x,y$ plane.  This fact may be confirmed by computing the moments
utilizing the covariant position operators. [See Ref.~\cite{park} for
the detailed computations of vortex positions.]  The energy of the
configuration does not depend on the transverse positions
$\varphi_l^I$ of D0 branes. Thus, the D0 branes are not the
deformation of D2 branes but independent entities of the theory. This
shows that the noncommutative gauge theory remembers the underlying
matrix theory and so is somewhat bigger than the theory of D2 branes.
 
We did analyze the small fluctuations 
around the above solution and found that there are $N\times M$ tachyonic 
modes among the off-diagonal fluctuations and their masses are 
\beq 
m^2_{l\alpha}= -\frac{1}{\theta}(1 - \sum_I(\varphi_l^I-h^I_\alpha)^2
) \, ,
\eeq 
where $l$ and $\alpha$  are respectively the indices for $M$ D0 and $N$ D2.
This result confirms that our solution indeed describes $M$ D0 branes 
near $N$ D2 branes. Using the unitary transformation,
\beq 
U= \sum_{l=0}^{M-1} |l \rangle   {\langle}l,0| + \sum_{p} |pN +M \rangle   
 {\langle}p+M,0| +  
\sum_{\alpha\neq 0,p} |pN+\alpha+M \rangle   {\langle}p,\alpha| \, ,
\eeq 
we can put all $D0$ in the first D2 brane, or $\alpha=0$ brane, which 
is closer to the $U(N)$ theory point of view.

A more nontrivial solution is the $U(N)$ dyonic fluxon solution in
three dimensional space, generalizing that of Gross and
Nekrasov\cite{grossb}.  Here the first two spaces $x,y$ are
noncommutative and the third coordinate $z$ is commutative. The
universality works in this higher dimensional theory which has six
adjoint scalar fields. This can be interpreted as the tilted $M$ D
string piercing the $N$ D3 branes, with some number of the fundamental
string bound to D3 branes. The $U(1)$ theory has a solution of two BPS
Dirac monopoles of opposite charge with the Dirac strings coming out
from the monopoles in the opposite directions. The fluxons are the
configurations where two BPS magnetic monopoles overlap exactly and so
their magnetic charges cancel each other.  With a single Higgs
$\phi=\phi^1$, the 1/2 BPS equations are
\beqn 
E_i = \sin\xi \nabla_i \phi^1\,, \ \ \  
B_i = \cos\xi \nabla_i \phi^1 \, ,
\eeqn 
where $i=1,2,3$. 
The dyonic fluxon solution  which describes the $M$ D strings plus some F 
string  piercing $N$ D3 branes is  
\beqn 
&& C = \sum_l  \lambda_l |l \rangle   {\langle}l| + \sum_{\alpha,p} \sqrt{p+1} 
|pN+M+\alpha \rangle   {\langle}(p+1)N+M+\alpha| \, , \\  
&& \phi^1 = \sum_{l=0}^{M-1} (\varphi_l - \frac{z}{\theta\cos\xi})|l  
\rangle   {\langle}l| 
 + \sum_{\alpha,p} h_\alpha |pN+ M+\alpha \rangle   {\langle}pN+M+\alpha|\,,  
\eeqn 
with $A_0 = - \phi^1  \sin \xi $ and  $A_{3}=0$. 
The magnetic field is then 
\beqn 
 B_3 = -\frac{1}{\theta} \sum_{l=0}^{M-1} |l \rangle   {\langle}l|  \, ,
\eeqn 
and the electric field $E_3= B_3 \tan\xi$.  The above solution is 
again given in the matrix theory point of view.  From the D3 world 
volume point of view, we can make the similar gauge transformation as 
for the D2-D0 case. The values $h_\alpha$ denote the $\alpha$-th D3 
brane position in the transverse $x^4$ direction.  From the above 
solution, we see that the $\l$-th fluxon is crossing the $\alpha$-th 
$D3$ brane at $z=\theta\cos\xi (\varphi_l -h_\alpha)$.

Five additional Higgs fields $\phi^J$ with $J=2,3,..,6$ can take 
expectation value without breaking the 1/2 BPS condition. 
Their energy contribution should be zero. The most general solutions 
are constant fields, 
\beq 
\phi^J = \sum_{l=0}^{M-1} \varphi_l^J |l \rangle    
{\langle}l| + \sum_{\alpha,p} h^J_\alpha 
|pN + M+\alpha  \rangle   {\langle}pN+M+\alpha| \, .
\eeq 
The value $\varphi^J$ denotes the position of the $l$-th fluxon along 
the transverse coordinate $x^{3+J}$ and the value $h_\alpha^J$ denotes 
the position of the $\alpha$-th D3 brane along the same transverse coordinate. 
Now there is an interesting possibility that the $l$-th fluxon may  
not meet any of D3 branes\cite{terashima},  
or may meet only a few of them.

While it is simple to see whether the $l$-th fluxon will meet the 
$\alpha$-th D3 branes from comparing their transverse positions, it 
will be more comforting to see this is the case by investigating the 
zero modes of the solution. When a fluxon goes through two parallel 
D3 branes, the middle segment can get separated from the whole line 
without change of energy. Thus, the number of the zero modes will 
depend on the number of such separable segments.

For the simplest case with a single fluxon and $\phi^J=0$, the
solution can be regarded as a composite of $N+1 $ magnetic monopoles,
related to the $U(N+2)$ gauge theory broken to $U(1)^{N+2}$, or the D
string connecting $N+1$ D3 branes. The first and the last D3 branes
are removed to the infinity and so the first and last segment of D
strings become infinitely long or their corresponding magnetic
monopoles become infinitely heavy and become Dirac monopoles with
finite tension Dirac strings on noncommutative space. The rest $N-1$ D
string segments have finite mass and appear as finite length sticks
whose ends appear as magnetic monopoles on corresponding $U(1)$ group.
The $U(N)$ fluxon solution is then one where all $U(N)$ magnetic
charges cancel each other exactly.  There would be then $4(N-1)$ zero
modes, four for each magnetic monopoles.  It would be the moduli space
of $N+1$ distinct magnetic monopoles such that the first and the last
one have infinite mass. For the fluxon case, the position of these two
infinitely massive monopoles would be identical. The metric of the
moduli space for the zero modes would be given by the Lee-Weinberg-Yi
type\cite{weinberg,klpy}.
 
Interesting generalization of these fluxon solutions would be 1/4 BPS 
and non BPS type of solutions.

Note added: After this paper was posted to hep-th,  we became aware of
the previous introduction of the covariant position on noncommutative
space in Ref.~\cite{stefan}, where some aspects of the issue was
discussed.

{\bf Acknowledgment} 
JHP would like to thank S. Hyun for the enlightening discussions.     
DB and KL are supported in part by KOSEF 1998 Interdisciplinary 
Research Grant 98-07-02-07-01-5.


\begin{thebibliography}{99} 
 
 
\bibitem{aschwarz} 
N. Nekrasov, and 
A.  Schwarz, 
Commun. Math. Phys. 198 (1998) 689, 
hep-th/9802068;  
K. Furuuchi, Prog. Theor. Phys. 103 (2000) 1043, 
hep-th/9912047.

\bibitem{klpy}
K. Lee and P. Yi,  
Phys. Rev. D61 (2000) 125015,  
hep-th/9911186. 
 
 
\bibitem{iso} 
N. Ishibashi, S.  
Iso, 
H. Kawai 
and Y. Kitazawa, 
Nucl. Phys. B573 (2000) 573,  
hep-th/9910004.  
 
\bibitem{itzhaki} 
D. J. Gross and A. Hashimoto,  
N. Itzhaki, 
hep-th/0008075.  
 
 
\bibitem{nekrasova} 
D. J. Gross and N. A. Nekrasov, 
hep-th/0010090.  
 
 
 
 
 
 
\bibitem{strominger}  
R. Gopakumar, S. Minwalla and A. Strominger, 
JHEP  05 (2000) 020,  
hep-th/0003160;  K. dasgupta, S. Mukhi, and G. Rajesh,
{\it Noncommutative Tachyons}, JHEP 0006 (2000) 022, hep-th/0005006;
J. A. Harvey, P. Kraus, F. Larsen, and E. J. Martinec, 
JHEP 07 (2000) 042,  
hep-th 0005031.
 
 
 
 
\bibitem{ahashimoto}  
A. Hashimoto and K. Hashimoto,  
JHEP 11 (1999) 005, 
{hep-th/9909202};  
%
%
%
 D. Bak,  
Phys. Lett. {B 471} (1999) 149, 
 hep-th/9910135;  
%
%
K. Hashimoto, H. Hata, and S. Moriyama,  
JHEP  12 (1999) 021,  
hep-th/9910196;  
D. Gross and N. Nekrasov,  
JHEP 07 (2000) 034,  
hep-th/0005204. 
 
 
\bibitem{grossb} 
D. J. Gross and N. A. Nekrasov,  
JHEP 10 (2000) 021,  
hep-th/0007204. 
 
 
 
  
 
\bibitem{klee} 
D. Bak and K. Lee, 
 hep-th/0007107. 
 
 
\bibitem{dbak} 
D. P. Jatkar, G. Mandal and S. R. Wadia,  
 JHEP 09 (2000) 018,  
hep-th/0007078. 
 
 
 


 
 
\bibitem{nsw} N. Seiberg and E. Witten,  
JHEP 09 (1999) 032,  hep-th/9908142.  
 

\bibitem{tfss} T. Banks, W. Fischler, S. H. Shenker and L. Susskind,  
Phys. Rev. D55 (1997)
5112, hep-th/9610043.




\bibitem{grt} O. J. Ganor, S. Ramgoolam and  W.  Taylor, 
Nucl. Phys. B492 (1997) 191,
hep-th/9611202; T. Banks, N. Seiberg and S. Shenker, 
 Nucl. Phys. B490 (1997) 91, hep-th/9612157.


 
\bibitem{aganagic} 
M. Aganagic, R. Gopakumar, S. Minwalla and  A. Strominger, 
 hep-th/0009142.  
 
\bibitem{polychronakos} 
A. P. Polychronakos,  
 hep-th/0007043; 
J. A. Harvey, P. Kraus and F. Larsen, 
hep-th/0010060; 
 K. Hashimoto, 
hep-th/0010251.  
%
  
 
 
 
 
 
 
\bibitem{park} 
D. Bak, 
hep-th/0008204; 
D. Bak, K. Lee, and J.-H. Park,  
hep-th/0011099.  
 
 
 
\bibitem{terashima} 
 M. Hamanaka and  S. Terashima, 
 hep-th/0010221.  
 
 
 
 
 
 
 
 
 
 
 
 
 
 
  
 
 
 
 
 
 
\bibitem{weinberg} 
K. Lee, E. J. Weinberg,  
and P. Yi,   
Phy. Rev. D54 (1996) 1633. 

 
 
 
\bibitem{stefan} J. Madore, S. Schraml, P. Schupp and J. Wess, Eur. Phys. J. 
C16 (2000) 161, hepth/0001203; B. Jurco, S. Schraml, P. Schupp and
J. Wess, Eur. Phys. J. C17 (2000) 521, hepth/0006246. 
 
 
 
 
 
 
 
 
 
 
\end{thebibliography}
\end{document}